\documentclass[oribibl]{llncs}

\PassOptionsToPackage{hyphens}{url}
\usepackage[colorlinks = true,
  linkcolor = blue,
  citecolor = blue,
  urlcolor = blue]{hyperref}

\usepackage{url}
\usepackage{csquotes}
\usepackage{multirow}
\usepackage{amsmath}
\usepackage{graphicx}
\usepackage{tabularx}
\usepackage{booktabs}
\usepackage{threeparttable}

\newcommand{\red}[1]{\textcolor{red}{#1}}

\usepackage{fancyhdr} 
\fancypagestyle{firststyle}
{
\fancyhf{}
\fancyfoot[C]{\scriptsize{Proceedings of the 20th International Conference on Availability, Reliability and Security (ARES~2025) Workshops, Ghent, Springer, pp. 269--285. This version is the authors' copy. The publisher's definite version is available online via \url{https://doi.org/10.1007/978-3-032-00633-2_16}.}}
}

\begin{document}

\title{A Time Series Analysis of Malware Uploads to Programming Language
  Ecosystems}

\author{Jukka Ruohonen\orcidID{\scriptsize{0000-0001-5147-3084}} \and \\ Mubashrah
  Saddiqa\orcidID{0000-0002-4816-2426} \\ \email{\{juk,
    msad\}@mmmi.sdu.dk}} \institute{University of Southern Denmark,
  S\o{}nderborg, Denmark}

\maketitle

\begin{abstract}
Software ecosystems built around programming languages have greatly facilitated
software development. At the same time, their security has increasingly been
acknowledged as a problem. To this end, the paper examines the previously
overlooked longitudinal aspects of software ecosystem security, focusing on
malware uploaded to six popular programming language ecosystems. The dataset
examined is based on the new Open Source Vulnerabilities (OSV)
database. According to the results, records about detected malware uploads in
the database have recently surpassed those addressing vulnerabilities in
packages distributed in the ecosystems. In the early 2025 even up to 80\% of all
entries in the OSV have been about malware. Regarding time series analysis of
malware frequencies and their shares to all database entries, good predictions
are available already by relatively simple autoregressive models using the
numbers of ecosystems, security advisories, and media and other articles as
predictors. With these results and the accompanying discussion, the paper
improves and advances the understanding of the thus far overlooked longitudinal
aspects of ecosystems and~malware.
\end{abstract}


\section{Introduction}

\thispagestyle{firststyle} 

Software ecosystems---understood in the present context as programming language
specific repositories from which software packages can be downloaded and
updated---have greatly facilitated software development and the general software
design principles, among them particularly reusability~\cite{Cox25}. This
facilitation has correlated with an enormous growth of the ecosystems, many of
which contain hundreds of thousands of software packages. However, over again,
software ecosystems have also been shown to be risky in terms of software
security. Indeed, the security aspects of practically all major software
ecosystems have been examined in recent years. The examples include, but are not
limited to, PyPI for Python~\cite{Guo23, Mehedi25, Ruohonen21PST, Zhang24}, CRAN
for R~\cite{Decan17}, npm for JavaScript~\cite{Zhang24, Zimermann19}, Maven for
Java~\cite{Nachuma25}, RubyGems for Ruby~\cite{Decan17}, and Packagist for
PHP~\cite{Ruohonen25CSA}. The overall conclusion from this already vast but
still growing literature base is the general insecurity of the software
ecosystems for individual developers and software development organizations,
whether companies or open source software projects.

Among the primary reasons for the security risks is that many of the packages
distributed in the ecosystems are of poor quality, containing various unverified
security issues or already identified vulnerabilities. The security risks also
increase due to a heavy use of dependencies in the ecosystems~\cite{Decan17,
  Nachuma25, Zimermann19}. Both direct and transitive dependencies contribute to
the risks, which are also related to the presence of many outdated,
unmaintained, and abandoned packages distributed in the ecosystems. Also the
operational security of software developers using the ecosystems has been seen
as a risk factor~\cite{Zhang24, Zimermann19}. Furthermore, the problems are made
worse by the increasing presence of malware in some programming language
ecosystems~\text{\cite{Guo23, Mehedi25, Zhang24}}. The paper aligns with and
contributes to the last mentioned genre of software ecosystem research, the
empirically motivated malware-specific research branch.

A traditional attack vector with the malware uploads has been so-called
typo-squatting; an attacker uploads a malware package with a name resembling an
existing, legitimate package, trying to fool people into downloading and
installing the malware-ridden package~\cite{Cao22}. Such typo-squatting belongs
to a broader class of name confusion attacks~\cite{snyk25}. For instance, it has
recently been argued that hallucinated package names by large language models
might make the problem worse in the nearby future~\cite{Register25}. Regardless,
the problem is already intensified by dependencies because it essentially may
only take one misstep by one developer somewhere in a dependency network to
compromise the whole network~\cite{Mehedi25}. For this reason alone, it is
important to gain better knowledge on the longitudinal aspects of malware
uploads to popular programming language ecosystems. With this point in mind, the
following three research questions (RQs) are examined:
\begin{description}
\itemsep 5pt
\item{RQ.1: How much detected malware uploads have popular programming language
  ecosystems seen compared to traditional vulnerability reports?}
\item{RQ.2: Which ecosystems have been particularly prone to malware uploads?}
\item{RQ.3: Can time series analysis provide insights into malware uploading
  trends?}
\end{description}

To the best of the authors' knowledge, the RQs, or something analogous, have not
been asked and examined in previous research. The paper thus fills a small but
notable knowledge gap. The first RQ.1 also reflects a recently proposed
distinction between supply chain attacks and vulnerabilities; the former can be
seen to involve also an insertion of malware into an ecosystem, whereas the
latter is about vulnerabilities in third-party components propagating into a
software using the components~\cite{Cox25}. Though, as was noted, also malware
may propagate through dependencies. If such a propagation reaches commercial
software vendors or important open source software projects, including
distributors such as Linux distributions, the consequences can be catastrophic
in many ways.

\section{Motivation}

The security risks involved are easy to demonstrate. Most---if not all---malware
recently discovered from the npm ecosystem come with the following warning:
\begin{displayquote}
``\textit{Any computer that has this package installed or running should be
    considered fully compromised. All secrets and keys stored on that computer
    should be rotated immediately from a different computer. The package should
    be removed, but as full control of the computer may have been given to an
    outside entity, there is no guarantee that removing the package will remove
    all malicious software resulting from installing
    it.''\footnote{~\url{https://osv.dev/vulnerability/MAL-2024-226}} } \\
\end{displayquote}

This serious warning is not intensified only by the noted risk with
dependencies. Among other things, many malware-ridden packages in both the npm
and PyPI ecosystems have relied on installation-time infections made possible by
the execution of scripts during installation or even
downloading~\cite{Zhang24}. Many of the malware uploads recorded in the OSV are
further referenced with CWE-506, which refers to embedding of malicious code in
the Common Weakness Enumeration (CWE) framework. A similar warning is available
from this framework.

The many security risks have been recognized also by policy-makers in recent
years. In terms of new regulations, particularly important to acknowledge is the
Cyber Resilience Act (CRA) recently enacted in the European Union
(EU)~\cite{EU24a}. This regulation is noteworthy for motivating also the present
work. Among other things, the CRA's new essential cyber security requirements
for most information technology products with a network functionality contain an
obligation to only ship products without known vulnerabilities. This requirement
applies also to vulnerabilities in dependencies distributed in software
ecosystems. Although the CRA does not mention malware explicitly, its further
essential requirements to ensure confidentiality, integrity, and availability
\cite{Ruohonen25RE} can be seen to cover also malware---as the above quotation
also testifies. Accidentally embedding malware to a product is thus likely to
face also regulatory sanctions at least in severe cases.

The CRA is important to mention also from a perspective of not software products
and producers but also from a perspective of regulators. In particular, the
regulation's Article~60 obliges European market surveillance authorities to
conduct coordinated sweeps of particular products for checking compliance and
detecting potential infringements.  Regarding the paper's time series analysis,
the note in the CRA's recital 114 about justifying sweeping particularly when
``market \textit{trends}, consumer complaints or \textit{other indications}
suggest that certain categories of products with digital elements are often
found to present cybersecurity \textit{risks}'' (italics added). While the CRA
is not meant to regulate all the world's software, sweeping software ecosystems,
possibly together with other stakeholders, might improve the cyber security for
everyone. To this end, it could be also argued that the new obligations placed
upon regulators themselves might enhance and improve the existing tracking and
monitoring infrastructures, among them the OSV database. After all, in the
context of cyber security, the concept of a sweep, which originates from the
EU's product safety laws~\cite{Chiara22}, is rather close to security scanning,
security audits, security monitoring, and related concepts and techniques
already used to improve also the security of software ecosystems.

\section{Related Work}

As already noted in the introduction, the paper's closest reference point is the
already large but still growing research domain on the security of programming
language ecosystems, including particularly its malware-motivated branch.

Given the time series analysis pursued, a few related works are worth remarking
also about longitudinal malware research---not least because such research seems
to be rather limited when compared to malware research in general. For instance,
a search from Scopus on 4 June 2025 with a keyword ``\textit{malware}'' yields
$28,890$ results when the search is restricted to papers' titles, abstracts, and
author-provided keywords. When a keyword ``\textit{time series}'' is added, only
$233$ results are returned. Although a definite conclusion is impossible to draw
from these numbers, it can be still concluded that time series analysis is a
nascent branch within the broader malware research domain. Adding a third
keyword ``\textit{autoregressive}'' yields only three papers. The first of these
is a paper for forecasting malware infection rates in higher education
institutions~\cite{deSouza24}. The second uses malware detection as a motivation
for an idea to design a new deep learning architecture for network traffic
analysis~\cite{Marwah22}. The third paper's topic is also about malware
detection but its methodology is based on an autoregressive moving average
model~\cite{Kim15}. These three studies signify what time series analysis
supposedly has often been about in the existing research: forecasts and
detection of malware, whether through outlier detection or by other
means. Although none of the three research questions explicitly fit into this
simple twofold categorization, particularly $\textmd{RQ}_3$ is close to the
first category because there would be only a small step from time series
model-building and trend analysis to forecasts.

\section{Materials and Methods}

\subsection{Data}\label{subsec: data}

The dataset examined was assembled from a bulk snapshot obtained in April 2025
from the OSV database.\footnote{~\url{https://osv.dev/}} The OSV database has
recently been used also in previous research~\cite{Ruohonen25ICTSS}. Although
the database curates data from various publicly available sources, the dataset
assembling was restricted to CRAN, Go, Maven, npm, PyPI, and RubyGems. Then, the
following time series were constructed:
\vspace{3pt}
\begin{enumerate}
\itemsep 5pt
\item{$\textit{MalFreq}_t$ counts the number of malware entries reported for all
  of the six ecosystems sampled at $t$. The identification of malware entries
  was done by including those files whose names started with a \texttt{MAL-}
  character string. Even though this simple identification technique is not
  perfect, searching for a string \texttt{malware} from the other files
  indicates no major concerns.}
\item{$\textit{MalShare}_t$ is a percentage share of malware entries to all
  entries in the six ecosystems at a given $t$. If $\textit{VulnFreq}_t$ would
  be a total count of software vulnerabilities in the six ecosystems at the
  given $t$, an approximation $\textit{MalShare}_t \simeq \textit{MalFreq}_t
  ~/~(\textit{MalFreq}_t + \textit{VulnFreq}_t) \times 100$ would hold.}
\item{$\textit{Eco}_t$ is a count of the given ecosystems that contributed to
  $\textit{MalFreq}_t$ at $t$. It follows that $\max(\textit{Eco}_t) = 6$ and
  $\min(\textit{Eco}_t) = 0$ hold for all $t$.}
\item{$\textit{Adv}_t$ is a count of security advisories curated in the OSV at
  $t$ for malware entries in the six ecosystems. Given the OSV's JavaScript
  Object Notation (JSON) schema, the parsing was done by searching and counting
  the \texttt{ADVISORY} entries in the schema's \texttt{references} field.}
\item{$\textit{Art}_t$ is a count of media articles, blog posts, and related
  information sources recorded in the OSV database at $t$ for malware entries in
  the six ecosystems. The parsing was analogous to $\textit{Adv}_t$ but by using
  the \texttt{ARTICLE} entries.}
\end{enumerate}
\vspace{3pt}

These five time series were operationalized into daily, weekly, and monthly
aggregates for which the lengths are $T = 1195$, $T = 168$, and $T = 39$,
respectively. The starting periods were restricted to the first day, first week,
and first month (January) of 2022. The reason for this restriction is that only
a few malware entries have been recorded in the OSV database prior to
2022. Regarding the daily aggregates, $\textit{MalShare}_t$ was manually set to
zero in case an amount of all entries at a day $t$ was zero. Given the date of
the data collection, the end periods are March 2025 and its last day and week.

\subsection{Methods}\label{subsec: methods}

The following autoregressive distributed lag (ARDL) model is used:
\begin{align}\label{eq: model}
  f(y_t) = \alpha &+ \sum^{p_1}_{j=1}\beta_j f(y_{t-j})
  + \sum^{p_2}_{j=0} \gamma_j f(\textit{Eco}_{t-j}) \\ \notag
  &+ \sum^{p_3}_{j=0} \phi_j f(\textit{Adv}_{t-j}) +
  \sum^{p_4}_{j=0}\rho_j f(\textit{Art}_{t-j}) + \varepsilon_t ,
\end{align}
where $y_t$ refers to either $\textit{MalFreq}_t$ or $\textit{MalShare}_t$, $t =
1, \ldots, T$, $\alpha$~is a constant, $\beta_j$, $\gamma_j$, $\phi_j$, and
$\rho_j$ are regression coefficients, and $\varepsilon_t$ is a normally
distributed residual term with a zero mean and a
variance~$\sigma^2_\varepsilon$. If $y_t$ is $\textit{MalFreq}_t$, \text{$f(x) =
  \ln(x + 1)$}; for $\textit{MalShare}_t$ it is an identity function, $f(x) =
x$. The immediate effects of a unit change in $\textit{Eco}_t$,
$\textit{Adv}_t$, and $\textit{Art}_t$ upon $y_t$ are given by $\gamma_0$,
$\phi_0$, and $\rho_0$, respectively. If the unit changes are sustained, the
effects are given by so-called long-run multipliers~(LRMs). To use
$\textit{Eco}_t$ as an example, such a multiplier is given by
\begin{equation}
\textmd{LRM}_{\textit{Eco}_t} =
\frac{\sum^{p_2}_{j=0} \gamma_j}{1 - \sum^{p_1}_{j=1}\beta_j} .
\end{equation}

The interpretation of these LRMs is similar to standard regression
coefficients. If $\textit{MalShare}_t$ and $\textit{Eco}_t$ are considered as an
example, a sustained increase by one ecosystem in the $\textit{Eco}_t$ series
will increase $\textit{MalShare}_t$ by $\textmd{LRM}_{\textit{Eco}_t}$
percentage points, all other things being constant. In addition, dynamic
multipliers (DMs) are useful for evaluating the dynamics of a given
effect~\cite{Cheng19, Menegaki19}. To again use $\textit{Eco}_t$ as an example,
for some integer $k > 1$ the DMs for it are given~by
\begin{equation}
(\textmd{DM}_{\textit{Eco}_t, 1}, \ldots,
\textmd{DM}_{\textit{Eco}_t, k})
 = \left(\frac{\partial f(y_{t})}{\partial\textit{Eco}_t}, \ldots,
    \frac{\partial f(y_{t+k})}{\partial\textit{Eco}_t}\right) , \quad t + k \leq T .
\end{equation}
For a sufficiently large $k$, it follows that
\begin{equation}
\sum^k_{i=1} \textmd{DM}_{\textit{Eco}_t, i} \simeq \textmd{LRM}_{\textit{Eco}_t} .
\end{equation}

Finally, there is the tricky problem of selecting the orders $p_1$, $p_2$,
$p_3$, and~$p_4$. On one hand, selecting too short orders may lead to the
omitted variable bias because relevant information is excluded. Too short orders
often lead to also other problems, including remaining autocorrelation in the
residual term $\varepsilon_t$. On the other hand, selecting too long orders
encounters the overfitting problem.

By inspecting automatic order selection algorithms in two different
implementations~\cite{ARDL, statsmodels}, it can be concluded that both
implementations yield extremely long orders both with the Akaike information
criterion (AIC) and the Bayesian information criterion (BIC). Therefore, a
manual but still systematic three-step procedure was used. In the first step the
orders were uniformly increased, $p_1 = p_2 = p_3 = p_4$, until no notable
autocorrelation was present in the residual terms. In the second step $p_1$, as
obtained from the first step, was held constant but the remaining orders, $p_2 =
p_3 = p_4$, were uniformly decreased until either autocorrelation was present or
any of the coefficients $\gamma_{p_2}$, $\phi_{p_3}$, $\rho_{p_4}$ were
statistically significant at the conventional 95\% confidence level. In the
third and final step $p_4$, $p_3$, and $p_2$, in the order of listing, were
consecutively and individually decreased by using the same stop criterion as in
the second step.

\section{Results}

\subsection{Descriptive Statistics}\label{subsec: descriptive statistics}

Th presentation of the results can be started by taking a look at the OSV's
malware entries across the six ecosystems; a basic breakdown is shown in
Table~\ref{tab: entries}. As can be seen, npm and PyPI have garnered the most
entries---as well as the most malware entries. In fact, as much as about 84\%
and 57\% of all entries for these two ecosystems have recently (from 2022
onward) been about malware. Although it is impossible to say how many have
fallen victim to these malware uploads, the observation is still quite alarming
in a sense that vetting of new uploads seems to be either working poorly or
absent altogether. Interestingly, furthermore, RubyGems takes only the fifth
place in terms of total entries but the third place in terms of malware
uploads. At the moment, CRAN, Go, and Maven seem to have not been particular
targets of malware uploads, or they have countermeasures in place, but the
situation may change in the future.

\begin{table*}[th!b]
\centering
\caption{Entries Across the Six Ecosystems}
\label{tab: entries}
\begin{tabular}{lcrcrcr}
  \toprule
&& \multicolumn{3}{c}{Frequency} && \multirow{2}{*}{Malware share} \\
\cmidrule{3-5}
Ecosystem &\qquad\qquad\qquad& All entries &\qquad\qquad& Malware entries &\qquad\qquad& \\
\hline
CRAN && $10$ && $0$ && $0.00$ \\
Go && $4,145$ && $8$ && $0.19$ \\
Maven && $5,461$ && $1$ && $0.02$ \\
npm && $24,837$ && $20,481$ && $82.46$ \\
PyPI && $15,929$ && $8,966$ && $56.29$ \\
RubyGems && $1,727$ && $813$ && $47.07$ \\
\bottomrule
\end{tabular}
\end{table*}

\begin{figure}[p!]
\centering
\includegraphics[width=\linewidth, height=6cm]{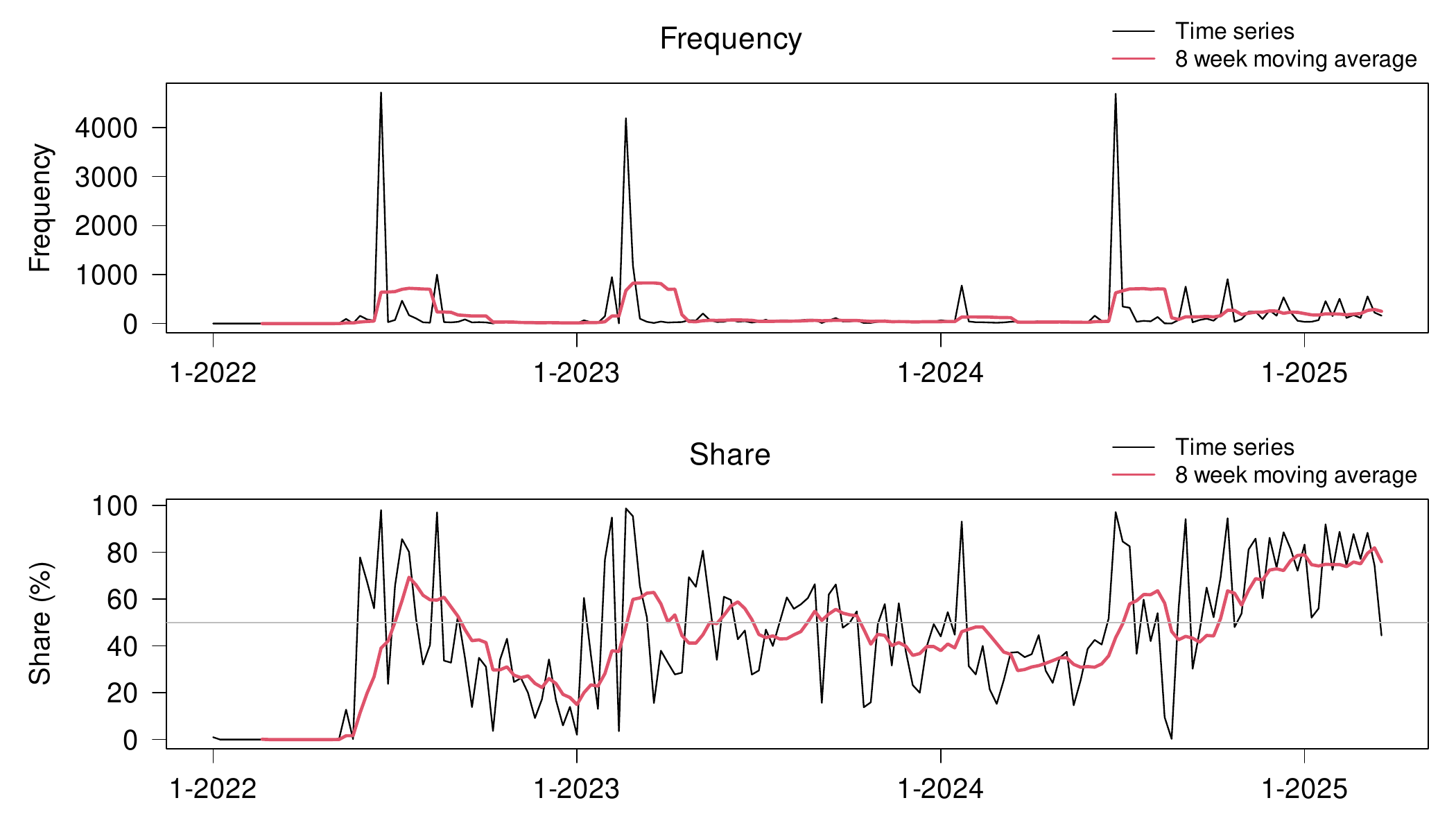}
\caption{The Two Malware Time Series (weekly aggregates)}
\label{fig: dependent series}
%
\vspace{30pt}
%
\centering
\includegraphics[width=\linewidth, height=9cm]{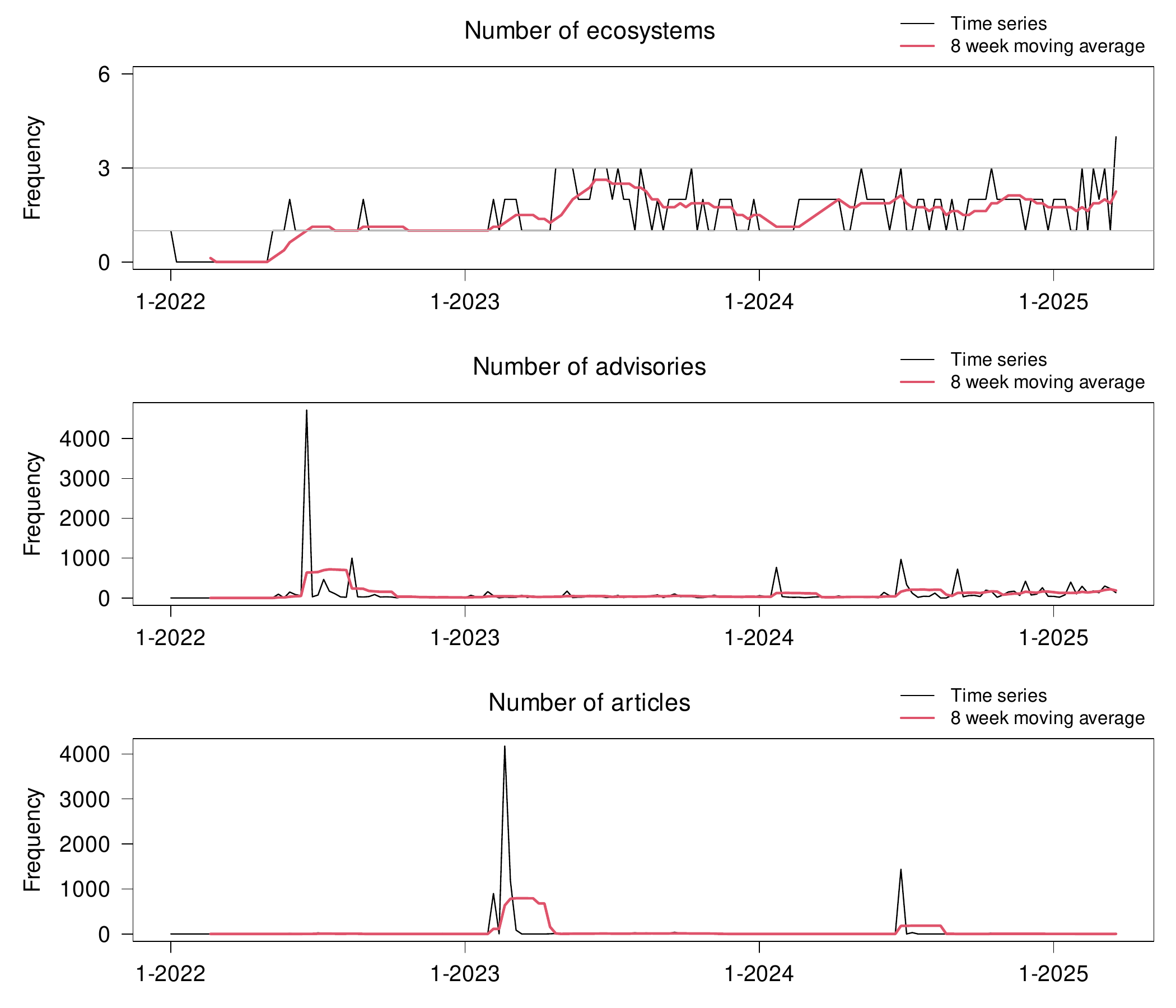}
\caption{The Three Explanatory Time Series (weekly aggregates)}
\label{fig: independent series}
\end{figure}

The weekly $\textit{MalFreq}_t$ and $\textit{MalShare}_t$ time series shown in
Fig.~\ref{fig: dependent series} further indicate the persistence of malware
uploads to the three ecosystems. The former series contain three large spikes,
which are likely due to specific sweeps or more general clean-up
operations. Even when keeping these spikes in mind, the median is as high as 37
malware entries per week, which is again a rather striking number on its
own. Then, the $\textit{MalShare}_t$ time series fluctuates a lot, partially due
to the spikes in $\textit{MalFreq}_t$ but also otherwise. This observation
reflects the operationalization of the time series (see Subsection~\ref{subsec:
  data}). The moving average shown was around 50\% during 2023 and most of 2024,
but in the early 2025 it had increased even up to 80\%. This amount is almost twice the median of 41\%.

Regarding the three explanatory time series shown in Fig.~\ref{fig: independent
  series}, $\textit{Eco}_t$ fluctuates mainly between the values one and three,
meaning that malware uploads into npm, PyPI, and RubyGems have sometimes been
reported individually in a given week but some other times weekly malware
reports have been made for all three ecosystems. However, a human eye cannot see
whether the fluctuations in $\textit{Eco}_t$ correspond with the fluctuations in
$\textit{MalFreq}_t$ and $\textit{MalShare}_t$. This point justifies the formal
ARDL modeling soon disseminated in Subsection~\ref{subsec: regression analysis}.

With respect to the two other explanatory time series, the three notable spikes
in $\textit{MalFreq}_t$ seem to correspond with one large spike in
$\textit{Adv}_t$ and two visible spikes in $\textit{Art}_t$. Also this
observation motivates the formal time series modeling because it seems that to
some extent publicity correlates with reported malware uploads. When taking a
peek at the sources behind the articles counted by $\textit{Art}_t$, many of
these have been either media articles and blogs about open source software
supply-chain security or malware discovery announcements from cyber security
companies and others scanning the programming language ecosystems. Given the
ARDL context, it can be hypothesized that publicity may not only correlate
simultaneously with the spikes but past publicity may influence a current or a
future discovery rate. If there is a lot of publicity, as has been the case in
the past three years, it may be that more and more companies, open source
software developers, cyber security professionals, and others pay attention to
the malware uploading problem. That is, it may be that $\textit{Adv}_{t-1},
\ldots, \textit{Adv}_{t-p_3}$ and $\textit{Art}_{t-1}, \ldots,
\textit{Art}_{t-p_4}$, not necessarily $\textit{Adv}_t$ and $\textit{Art}_t$
alone, influence $\textit{MalFreq}_t$ and $\textit{MalShare}_t$. A similar
reasoning applies to $\textit{Eco}_t$. The rising trends in $\textit{MalFreq}_t$
and $\textit{MalShare}_t$ from late 2024 onward, and the persistence of the $[1,
  3]$ range fluctuations in $\textit{Eco}_t$, \text{may---or}
\text{should---motivate} further security scans and sweeps---or, alternatively,
these should not at least motivate stopping existing efforts.

\subsection{Regression Analysis}\label{subsec: regression analysis}

\begin{figure}[p!]
\centering
\includegraphics[width=\linewidth, height=8cm]{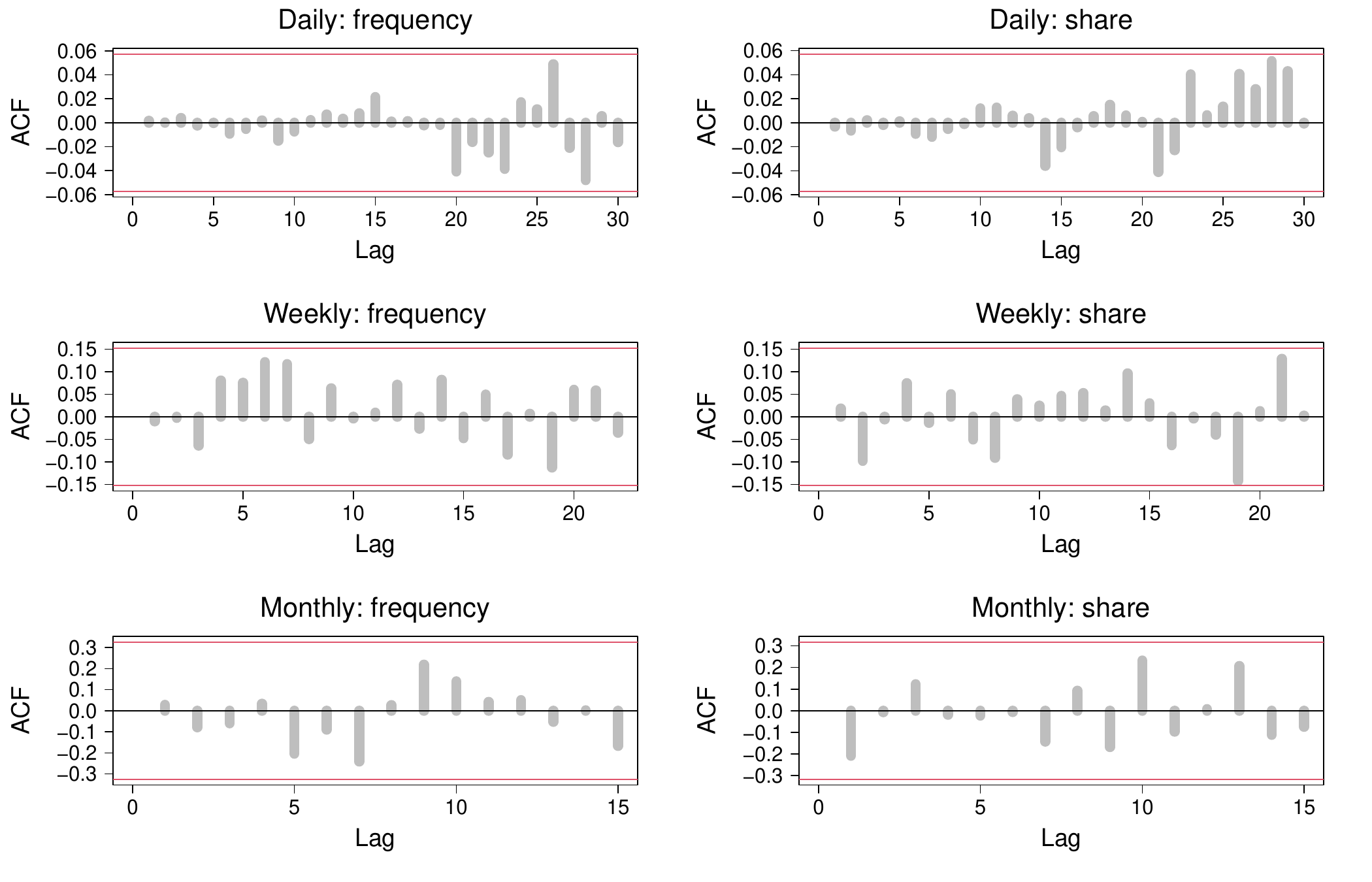}
\caption{Autocorrelation Functions of the Residual Terms from the Six ARDL Models (95\% confidence intervals; maximum lag lengths determined by $\lfloor 10 \times \log_{10}(T) \rfloor$)}
\label{fig: acf}
%
\vspace{30pt}
%
\centering
\includegraphics[width=\linewidth, height=8cm]{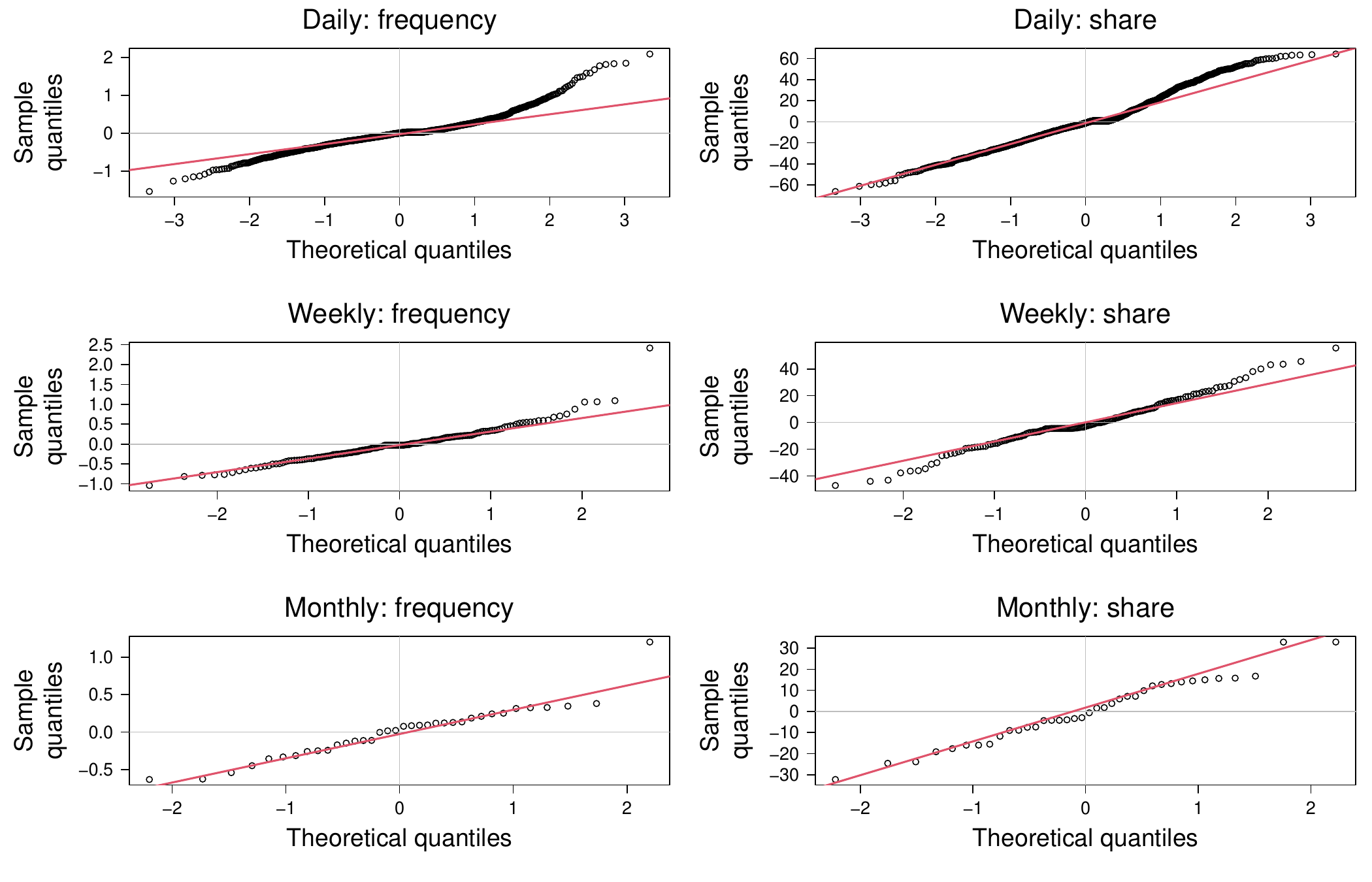}
\caption{Normal Quantile-Quantile Plots}
\label{fig: qq}
\end{figure}

The ARDL model in \eqref{eq: model} is estimated for both $\textit{MalFreq}_t$
and $\textit{MalShare}_t$ by using the daily, weekly, and monthly
aggregates. Thus, in total six models are estimated. Before continuing, it can
be noted that both series are stationary, as also confirmed by formal
Dickey-Fuller tests~\cite{Dickey79}. Another preliminary point is about the
manual order selection procedure described in Subsection~\ref{subsec:
  methods}. As can be seen from Table~\ref{tab: orders}, the procedure resulted
relatively, but not substantially, large orders for the daily aggregates, as
could be expected. Somewhat unexpectedly, however, rather short orders were
suitable for the weekly aggregates. A minimal model with $p_1 = 1$ and $p_2 =
p_3 = p_4 = 0$ was suitable for the monthly $\textit{MalShare}_t$
series. Despite these points, the procedure worked well in ensuring that no
remaining autocorrelation is present. As can be seen from Fig.~\ref{fig: acf},
all autocorrelation functions (ACFs) remain below the 95\% confidence
intervals.

\begin{table*}[th!b]
\centering
\caption{ARDL($p_1, p_2, p_3, p_4$) Orders Selected}
\label{tab: orders}
\begin{tabular}{lcccccc}
\toprule
&& Daily && Weekly && Monthly \\
\hline
Frequency &\qquad\qquad& (27, 26, 22, 18) &\qquad\qquad& (3, 2, 2, 2) &\qquad\qquad& (3, 2, 3, 2) \\
Share && (23, 21, 21, 20) && (3, 0, 1, 0) && (1, 0, 0, 0) \\
\bottomrule
\end{tabular}
\end{table*}

As is typical in applied problems, the normality assumptions about
$\varepsilon_t$ is a small problem. For instance, the Jarque-Bera
test~\cite{Jarque80}, which is based on skewness and kurtosis, rejects the null
hypothesis of normality for all residual series except those coming from the
weekly and monthly estimates for $\textit{MalShare}_t$. When taking a visual
look at Fig.~\ref{fig: qq}, however, the situation is hardly as bad as the
formal test results would indicate. In other words, all residual series resemble
the normal distribution; some more, some less, but all still sufficiently for
proceeding.

Heteroskedasticity is also a slight problem. As can be seen from Fig.~\ref{fig:
  er}, particularly the residuals from the two models for the daily aggregates
indicate non-random patterns. There is also a related problem: the plain ARDL
model is not optimal for $\textit{MalShare}_t$ because
$\max(\textit{MaxShare}_t) = 100$, which is exceeded by some of the estimated
values. However, such exceedances are rather small: the percentage shares of
estimated values exceeding one hundred are only $1.7$, $1.8$, and $2.6$ for the
three models using the daily, weekly, and monthly aggregates of
$\textit{MalShare}_t$, respectively. All in all, it must be acknowledged that
some diagnostic problems are present, as is often the case in applied time
series regression analysis, but none of the problems are severe enough to
prevent proceeding into the actual results. Against this backdrop, the LRMs are
shown in Table~\ref{tab: lrm}.

\begin{figure}[th!b]
\centering
\includegraphics[width=\linewidth, height=8cm]{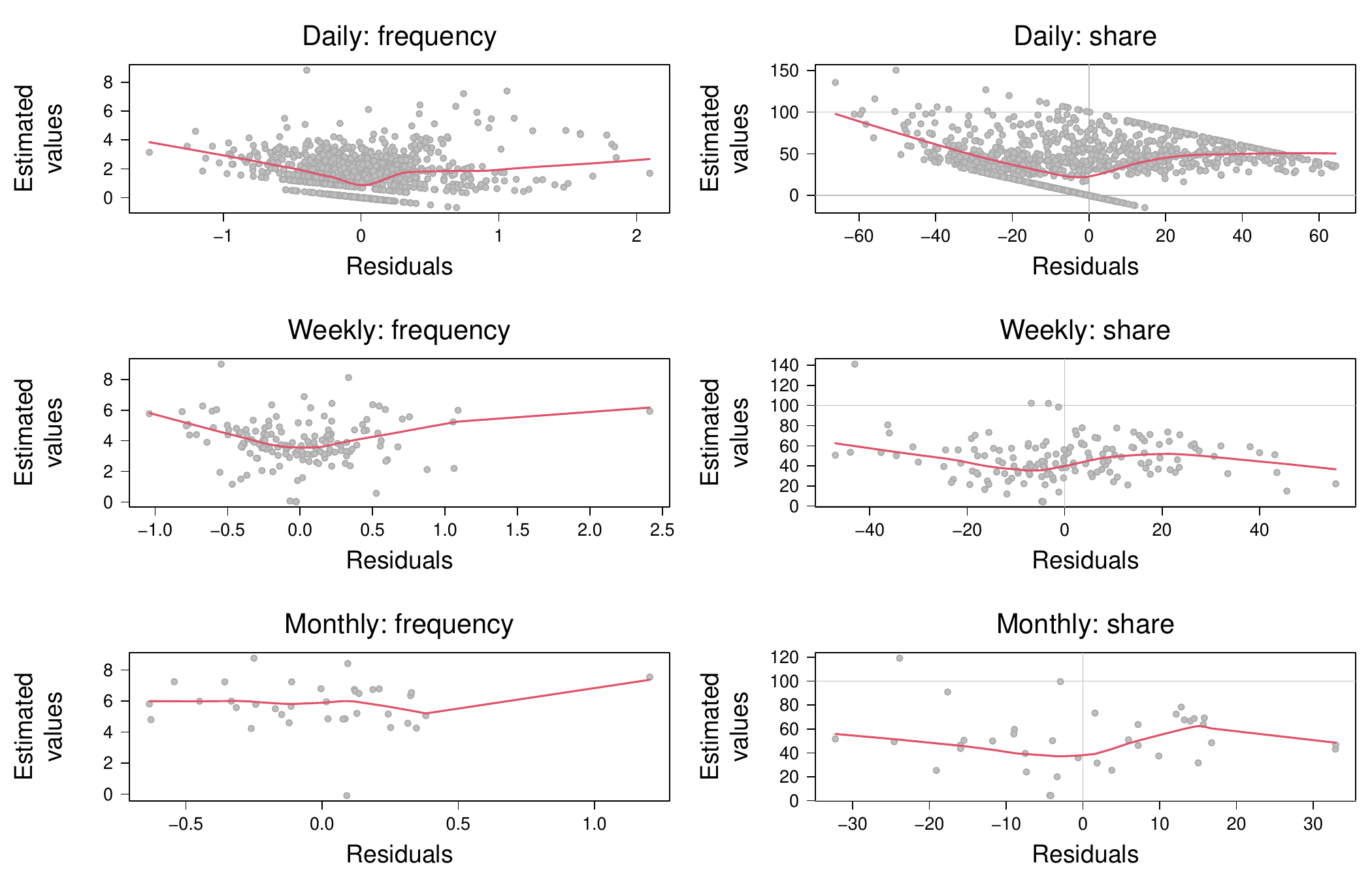}
\caption{Estimated Values and Residuals}
\label{fig: er}
\end{figure}

\begin{figure}[p!]
\centering
\includegraphics[width=\linewidth, height=18cm]{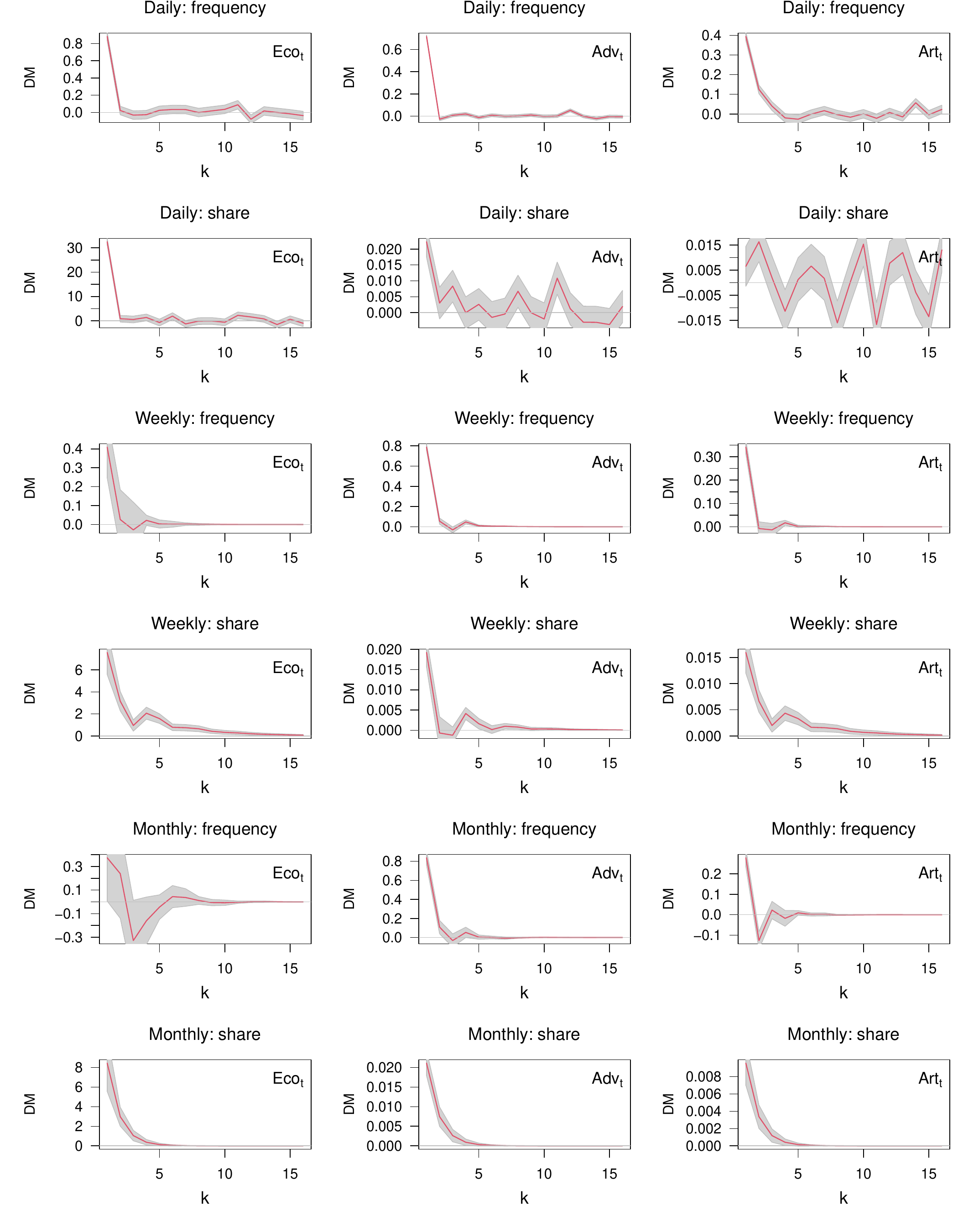}
\caption{Dynamic Multipliers}
\label{fig: dm}
\end{figure}

\begin{table*}[th!b]
\centering
\caption{Long-Run Multipliers$^1$}
\label{tab: lrm}
\begin{threeparttable}
\begin{tabular}{lcrcrcrcrcrcr}
\toprule
&& \multicolumn{3}{c}{Daily}
&& \multicolumn{3}{c}{Weekly}
&& \multicolumn{3}{c}{Monthly} \\
\cmidrule{3-5} \cmidrule{7-9} \cmidrule{11-13}
&\qquad\qquad& Frequency &\qquad& Share &\qquad\qquad& Frequency &\qquad& Share &\qquad\qquad& Frequency &\qquad& Share \\
\hline
$\textit{Eco}_t$ && \red{0.812} && \red{46.501} && 0.437 && \red{19.413} && 0.160 && \red{13.060} \\
$\textit{Adv}_t$ && \red{0.840} && \red{0.113} && \red{0.886} && \red{0.027} && \red{0.963} && \red{0.033} \\
$\textit{Art}_t$ && \red{0.519} && 0.061 && \red{0.344} && \red{0.041} && \red{0.169} && \red{0.015} \\
\bottomrule
\end{tabular}
\begin{tablenotes}
\begin{scriptsize}
\item{$^1$~Colored entries denote statistical significance at the conventional
  95\% level.}
\end{scriptsize}
\end{tablenotes}
\end{threeparttable}
\end{table*}

All three explanatory time series indicate long-run effects, irrespective
whether daily, weekly, or monthly aggregates are used. Furthermore, only three
of the LRMs are not statistically significant at the conventional 95\%
confidence level. Having said that, the effects from $\textit{Adv}_t$ and
$\textit{Art}_t$ are much lower than those from the ecosystem count time series,
which indicate substantial long-run impacts. All other things being constant, a
unit increase in $\textit{Eco}_t$ increases $\textit{MalShare}_t$ by $46.5$
percentage points daily, $19.4$ percentage points weekly, and $13.1$ percentage
points monthly. When keeping the maximum in mind, these long-run effects are
substantial but hardly surprising as such due to the concentration of malware
uploads to npm, PyPI, and RubyGems. The DMs shown in Fig.~\ref{fig: dm} further
indicate that the effects are not merely immediate shocks but persist relatively
long before eventually dampening. Though, a similar observation applies to the
other series as well. The $\textit{Adv}_t$ and $\textit{Art}_t$ time series
indicate particularly disturbing dynamic shocks upon $\textit{MalShare}_t$ in
the model using the daily aggregates. Although the corresponding effects, as
seen from the $y$-axes in Fig.~\ref{fig: dm} and the LRMs in Table~\ref{tab:
  lrm}, are still small in magnitude, these persistent but fluctuating shocks
could be interpreted to support a conclusion that the publicity theorization is
not entirely without a basis. As for the earlier speculation in the previous
Subsection~\ref{subsec: descriptive statistics} about the potential impact of
the past values of the ecosystem time series, it can be concluded
\text{that---and} despite for the selection of $p_2 = 0$ for two series (see
\text{Table~\ref{tab: orders})---the} effects are not entirely
simultaneous. Finally, the empirical exposition can be ended by noting that the
ARDL models yield generally good statistical performance; the lowest and highest
coefficients of determination are $0.59$ and $0.95$. While forecasts can be left
for further work, such values indicate that even simple models could be used
also in practical foresight.

\section{Discussion}

\subsection{Conclusion}

The paper examined recent (from 2022 to early 2025) malware uploads to six
popular programming language ecosystems: CRAN, Go, Maven, npm, PyPI, and
RubyGems. Regarding the three research questions specified, the answer to RQ.1
is simple but alarming: malware uploads have surpassed the reporting of
traditional software vulnerabilities in packages distributed in the
ecosystems. With respect to RQ.2, npm (over twenty thousand malware uploads),
PyPI (nearly nine thousand malware uploads), and RubyGems (about eight hundred
malware uploads) have been particularly prone to malware uploads, whereas CRAN,
Go, and Maven have seen less than ten malware uploads in total. The answer to
the third and final RQ.3 is that time series analysis can reveal insights about
malware uploads and their trends. The decent statistical performance
\text{obtained---the} average coefficient of determination is
\text{$0.79$---indicates} that forecasting could be used also in practical
foresight; a hypothesis is that the increasing trend of malware uploads
continues also in the nearby future. As could be expected, the number of
ecosystems provides particularly good predictive power; the more there are
ecosystems, the more malware uploads are also reported. Smaller but still
visible effects are present for security advisories and media and other
articles; publicity seems to also affect the malware upload trends. Rather
analogously to recent arguments about reported vulnerabilities in open source
software projects~\cite{Ruohonen25CSA}, the explanation might be that increasing
publicity about malware uploads prompts more companies and security
professionals to scan and monitor the ecosystems.

\subsection{Limitations}

A couple of noteworthy limitations can be acknowledged. The first is that the
OSV database has not been yet validated in research. Given the well-known
problems with other vulnerability databases~\cite{Anwar22, Esposito23,
  Massacci13}, it may be that also the OSV database suffers from different
reliability and validity problems. These problems cover also external validity;
it is not clear how good is the database's coverage in terms of the individual
tracking databases used by open source software projects.

The second limitation is more theoretical: only known and reported malware cases
were observed. Rather analogously to observing only known and reported
vulnerabilities~\cite{Ruohonen25CSA}, nothing can be deduced about what a
theoretical malware count, say $\textit{TheoMalFreq}_t$, might be. If
$\textit{TheoMalFreq}_t$ would include both $\textit{MalFreq}_t$ and cases that
are unknown but still true positives, $\textit{TheoMalFreq}_t \geq
\textit{MalFreq}_t$ should hold yet $\sigma = \textit{TheoMalFreq}_t -
\textit{MalFreq}_t$ remains unknown. This point is important because a risk
analysis should assess also $\sigma$. Given the sizes of the ecosystems
observed, nevertheless, it can tentatively concluded that an unconditional
probability of picking a malware package is supposedly still tiny---even when
dependencies are accounted for. However, further research is needed about the
effectiveness of typo-squatting. Analogously to phishing and
scamming~\cite{Ma25, Ruohonen24ICDF2C}, the effectiveness of typo-squatting
cannot be fully understood by quantitative security metrics alone; psychology
and other factors must be accounted too.

\subsection{Implications}

The research on ecosystems and malware has often recommended improving
monitoring and detection capabilities~\cite{Cao22, Guo23}. In addition to
publicity and awareness, it may be that this recommendation also implicitly and
partially explains the answer to RQ.2. In other words, it may---or may
not~\cite{Goodin25}---be that detection and monitoring capabilities might have
already improved, such that more malware uploads have been detected, removed,
and \text{reported---possibly} irrespective whether malware uploads have
actually increased \textit{per~se}. While improving the capabilities further may
improve the situation somewhat, a probability of bad apples slipping through is
also dependent on the sizes of the ecosystems. When there are hundreds of
thousands of packages, it is probable that some malware will slip through even
with highly accurate detection engines. Against this backdrop, it is interesting
to see whether the future will see curated lists for safe and secure packages. A
recommendation to improve code signing~\cite{Cox25} aligns with such curated
lists. Curating has also been what Linux distributions have always done, but
somewhere in recent history this quality gating function was forgotten or
overridden by the emergence of programming language software ecosystems.

The results have also implications for research. For instance, reflecting a lack
of data and code sharing~\cite{SEI25, Zheng18}, as well as a lack of good
benchmark datasets~\cite{Botacin21, Zahan24}, some studies have attempted to
verify a true positiveness of a malware sample by checking that the malware
packages are absent in PyPI~\cite{Mehedi25}. Clearly, such a check is misleading
or at least a poor choice due to the prevalence of malware in PyPI too. Instead,
a starting point for further research might be to evaluate the performance of
existing commercial malware detection engines in the programming ecosystem
context.\footnote{~\url{https://www.virustotal.com/}} It may be that detection
accuracy is not as good as with other, more conventional malware variants
usually distributed as binaries. There is also room for more practice-oriented
work regarding takedown efficiency, which seems to be suboptimal according to
existing research~\cite{Cao22}. Though, takedowns and clean-ups reiterate also
the point about curated lists and signing---it is debatable to which areas
future efforts should be allocated. Finally, the earlier point about the
effectiveness of typo-squatting should be evaluated also against the new issues
brought by large language models~\cite{Register25, Luo25}. By hypothesis,
susceptibility to typo-squatting increases when using these~models.

\bibliographystyle{splncs03}

\end{document}